\documentstyle[twocolumn,aps,epsf]{revtex}

\def\myfigure#1#2{{\leftskip=0.000753\textwidth \rightskip\leftskip\small
\begin{figure}\baselineskip=14pt plus 2pt minus 1pt
\centerline{#1}\nobreak\smallskip\nobreak #2\end{figure}}}
\global\firstfigfalse

\draft	

\begin{document}
\title{First Passage Times and Time-Temperature-Transformation curves for
Martensites }

\author{Madan Rao$^{1}$ and Surajit Sengupta$^2$}

\address{$^1$Institute of Mathematical Sciences, Taramani, Madras 600113,
India\\ $^2$Material Science Division, Indira Gandhi Centre for Atomic
Research, Kalpakkam 603102, India}

\date{\today}

\maketitle

\begin{abstract}

Martensites are long-lived nonequilibrium structures produced
following a quench across a solid state structural transition.  In a
recent paper (Phys. Rev. Lett. {\bf 78}, 2168 (1997)), we had
described a mode-coupling theory for the morphology and nucleation
kinetics of the equilibrium ferrite phase and twinned martensites.
Here we calculate nucleation rates within a first-passage time
formalism, and derive the time-temperature-transformation (TTT)
diagram of the ferrite-martensite system for athermal and isothermal
martensites. Empirically obtained TTT curves are extensively used by
metallurgists to design heat treatment cycles in real materials. 

\end{abstract}


A martensitic transformation \cite{NISHROIT} is a nonequilibrium solid
state structural transition which results in a metastable phase ({\it
martensite}), often consisting of alternating twinned arrays within the
parent solid ({\it austenite}).  A characteristic of this
transformation is that it is diffusionless, i.e., the velocity of the
transformation front is much larger than typical diffusional speeds of
atoms. In contrast, infinitesimally slow cooling results in the
equilibrium phase ({\it ferrite}). 

Martensites are commercially important, for example as the hard
constituent in steel, or in shape-memory alloys like Nitinol. These
desirable properties, depend on the amount of transformed martensite. Over
the years a tremendous amount of empirical \cite{NISHROIT} and theoretical
\cite{KRUM,GOOD} knowledge regarding the kinetics of this transformation
has accrued. A convenient representation, used extensively by
metallurgists is the time-temperature-transformation diagram \cite{MH}.
The TTT diagram (Fig.\ 1), is a family of curves parametrized by the
fraction $\delta$ of transformed product. Each curve is a plot of the time
required to obtain $\delta$ versus temperature of quench. It is clear from
the figure that the shape of the curves are qualitatively different for
the ferrite and martensite products.  There are two important features
regarding the TTT curves for martensites. ($i$) The transformation curves
lie below a temperature $M_{s}$, the martensite-start temperature. This
means that a martensite does not obtain at temperatures above $M_s$, even
if the quench rates are infinitely high. ($ii$) The curves are parallel to
the time axis. This implies that the martensitic transformation is
completed immediately following the quench (such martensites are called
{\it athermal} martensites). Such transformation curves have been obtained
empirically\,; in the absence of a unified nonequilibrium theory for
martensite and ferrite nucleation, these curves have not been `derived'. 

In a recent paper \cite{PRL} we presented a mode-coupling theory for the
nucleation and growth of a product crystalline droplet within a parent
crystal. We showed that for slow quenches, the droplet grows diffusively
as an equilibrium ferrite inclusion, while for fast quenches, the droplet
grows ballistically, as a martensite having twinned internal substructure,
with a speed comparable to the sound velocity. Given this unified
description, can we arrive at the general features of the
phenomenologically obtained TTT curves ? The details of the underlying
physics are ofcourse complex --- this includes an understanding of the
heterogeneous nucleation and growth of martensitic grains and other
intervening phases, and the subsequent collision of grains emanating from
correlated nucleation events. In this Letter, we demonstrate that certain
qualitative apsects of the TTT diagram can nevertheless be obtained using
very general features of the distribution of nucleation events and the
nucleation dynamics arising from our mode-coupling theory \cite{PRL}. 

\myfigure{\epsfysize2.2in\epsfbox{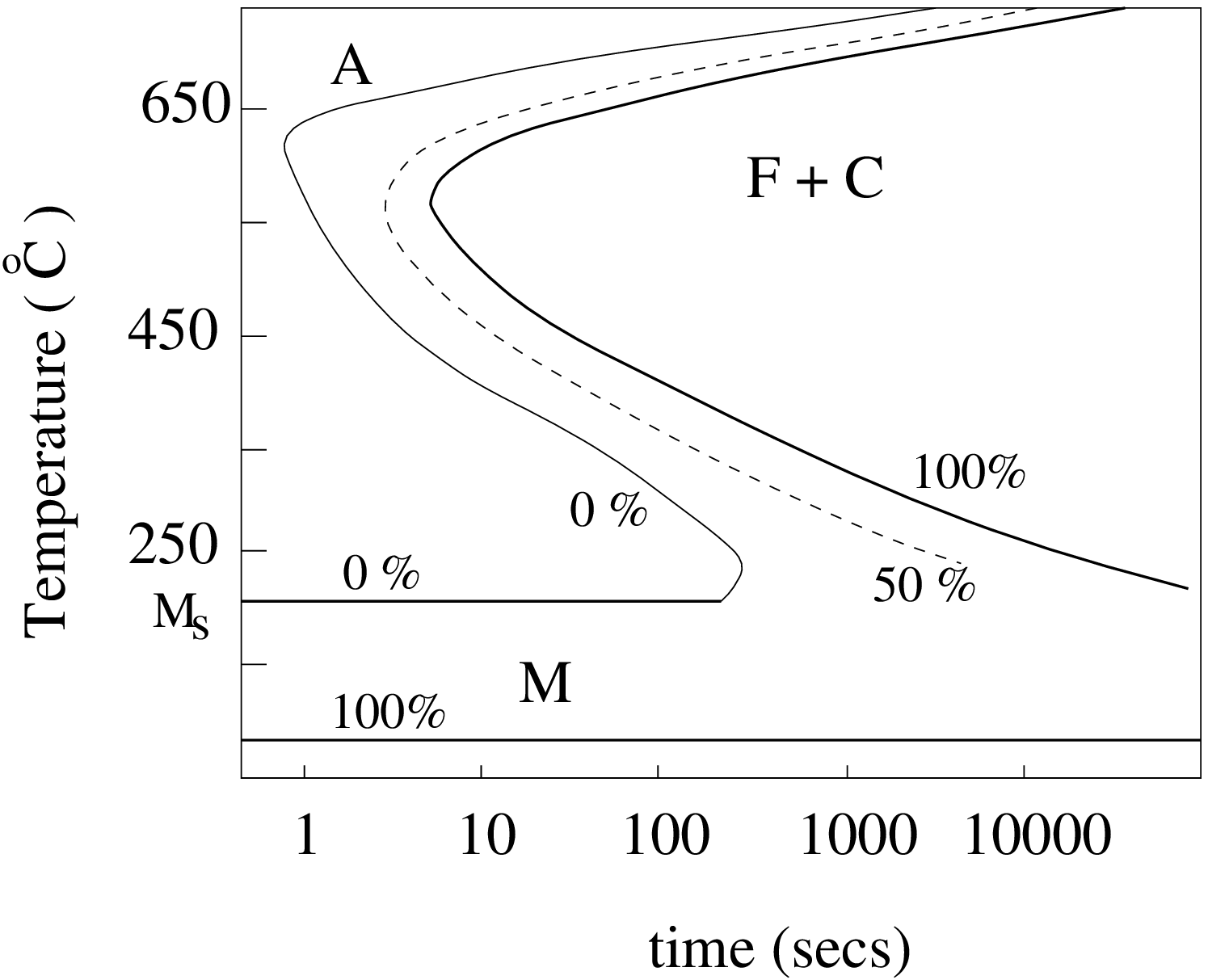}}{\vskip0inFig.\ 
1~~Experimental TTT curves (adapted from Ref. \cite{MH}) for the structural 
steel AISI 1090, containing approximately
$0.84 \%$ C and $0.60 \%$ Mn. The letters A, F and C represent
the f.c.c.-austenite, the equilibrium b.c.c-ferrite 
and the carbide precipitate (Fe$_3$C) respectively. 
Curves corresponding to $0$, $50$ and $100$ \% transformation are shown.
Below a temperature $M_s$, the metastable martensite (M) is formed - 
the transformation curves for martensites are horizontal.}

Though our analysis can be generalized to any structural transition in
arbitrary dimensions, we will, for calculational simplicity, focus on
the square to rhombus structural transition in 2-dimensions. It will
be clear that the qualitative form of the leading `edge' of the TTT
curves (labelled $0$\% in Fig.\ 1) is independent of dimensionality and 
the type of structural transition. 

The square to rhombus structural transformation involves a shear+volume
deformation, and so the strain order parameter $\epsilon_{ij}$ has only
one nontrivial component $\epsilon=(\partial_yu_{x}+\partial_xu_{y})/2$
(${\bf u}$ is the displacement vector field). A quench across this
transition results in the nucleation of a rhombic product in a square
parent. At the initial time, the transformed region is simply obtained
as a geometrical deformation of the parent creating a local atomic
mismatch \cite{NISHROIT}. This leads to a discontinuity in the normal
component of the displacement field across the parent-product interface
$\Delta {\bf u} \cdot {\hat {\bf n}}$ (${\hat {\bf n}}$ is the unit
normal to the interface) \cite{LL}. This discontinuity appears as the
vacancy field $\xi \phi \equiv \xi (n_{int} - n_{vac})/{\bar n}$, where
$\xi$ is the interfacial thickness, ${\bar n}$ is the average number
density, and $n_{int}$ and $n_{vac}$ are the interstitial and vacancy
densities respectively (measure of the compression or dilation of the
local atomic environment)\cite{NOTE1}. 

The free-energy functional ${\cal F}$ describing this inhomogeneous
configuration contains the usual bulk elastic free-energy of a solid
$F_{el}$ and an extra interfacial term describing the parent-product
interface \cite{PRL}. In our 2-d example, the bulk free energy $F_{el}$
is constructed to have three minima --- one corresponding to the
undeformed square cell ($\epsilon=0$) and the other two corresponding
to the two variants of the rhombic cell ($\epsilon=\pm \epsilon_{0}$).
The total (dimensionless) free-energy functional to leading order in
$\phi$ is,
\begin{equation} 
{\cal {F}} = \int_{x, y}
a\,{\epsilon}^2-{\epsilon}^4+{\epsilon}^6 +(\nabla \epsilon)^2
  + \frac {\gamma}{2} \,(\phi\, \partial_n \epsilon)^2 \xi^2\,\,.
\label{eq:FREE2D}
\end{equation}
The modulus $\gamma \equiv {\Omega_0}^{-1}\sum^{\prime} \xi^2
\partial_n \partial_n c(r)$ (prime denotes a sum across the interface)
is the surface compressibility of the vacancy field and depends
on the local orientation of the parent-product interface ($c(r)$ is
the direct correlation function of the liquid at freezing). The three
minima of the homogeneous part of $F$ at $\epsilon=0$ (square) and
$\epsilon \equiv \pm \epsilon_{0} = \pm [(1+\sqrt{1-3a})/3]^{1/2}$
(rhombus), are obtained in the parameter range $0 < a < 1/3$. The
first-order structural transition from the square to rhombus occurs at
$a=1/4$. For $1/4 > a > 0$, the square is metastable. The degree of
undercooling $1/4 - a \propto T_s - T$, where $T_s$ is the temperature
at which the equilibrium structural transition occurs. 

A quench across the structural transition, nucleates a region of the
product ($\epsilon=\pm \epsilon_0$) within the parent ($\epsilon = 0$). 
The growth of this nucleus is described by a Langevin equation for the
broken symmetry variable $\epsilon$ and the vacancy field $\phi$. Moving
with the growing interface, the equations of motion become,
\vspace {-.2 cm} 
\begin{equation} 
\frac {\partial \epsilon} {\partial t} =
-\,\Gamma \frac {\delta {\cal {F}}} {\delta \epsilon} + \eta
\label{eq:strain} 
\end{equation} 
\begin{equation} 
\frac {\partial \phi} {\partial t} = D_{\phi} {\nabla^2} \frac {\delta
{\cal {F}}} {\delta {\phi}} \,\,. 
\label{eq:phi} 
\end{equation} 
where the noise $\eta$ is an uncorrelated gaussian white noise with
variance proportional to temperature $\langle \eta(x,t) \eta(x',t')
\rangle = 2 k_BT \Gamma \delta(x-x') \delta(t-t')$. $D_{\phi}$ is the
microscopic vacancy diffusion coefficient which has 
an Arrhenius dependence $D_{\phi} =
D_{\infty} \exp (-A/k_B T)$. Typical values of parameters 
for pure ($99.98$ \%) Fe at around $1223 - 1473$ $^{\circ}$K have been 
measured to be\cite{SMITH} $A = 280-310$ kJ-mol$^{-1}$ and 
$D_{\infty} = 0.4-4.0$ cm$^2$/s, which fixes $D_{\phi} \sim 10^{-12}$
cm$^2$/s.

Our free-energy functional naturally admits two widely separated time
scales --- $\tau_l$, the relaxation time of the order parameter to the
local minima and $\tau_n$, the first-passage time for the order
parameter to go from the local to the global minimum. The shorter time
scale $\tau_l \sim \left[{\cal F}''(\epsilon=0) \right]^{-1/2} = \xi/c$
($c$ is the velocity of transverse sound) with a typical value of
$10^{-14}$s \cite{AS}. The relaxation time of $\phi$, given by
$\tau_{\phi} \sim \xi^2/D_{\phi}$, lies between $\tau_l \ll \tau_{\phi}
< \tau_n$. Thus the order parameter is `slaved' to $\phi$, which acts
as a time dependent source coupling to $\partial_n \epsilon$. The
`instantaneous' value of $\phi$ is given by the Eq.\ (\ref{eq:phi}). 

The exact formula for the nucleation rate of a solid nucleus growing
within a solid parent is unknown. In a continuum field theory
approximation, the first passage time $\tau_n$ would be given by the
Kramers formula with a time-dependent energy barrier \cite{AS}. In
this paper we shall take $\tau_{n} = \Gamma^{-1}\,\,\exp(\Delta E^*)$,
where $\Delta E^{*}$ is the energy barrier (in units of $k_BT$) to
form the critical nucleus. We have ignored the curvature dependent
prefactors since they are only weakly dependent on $\phi(t)$ and $a$. 

To calculate the energy of the critical nucleus, we use a 
variational ansatz for the strain profile defining a nucleus of linear 
dimension $L$,
\begin{equation}
\epsilon\,(x,y) = \left \{ \begin{array}{ll}
\epsilon_0   &  \mbox{if \,\,\,$ -L/2+\xi/2 < x < L/2 
-\xi/2$} \\
      &  \mbox{and \,$-L/2+\xi/2 < y < L/2 -\xi/2$} \\
      &      \\
0     &  \mbox{if \,\,\,$-L/2-\xi/2 > x > L/2+\xi/2$} \\
      &  \mbox{or \,\,\,$-L/2-\xi/2 > y > L/2+\xi/2$}
\end{array}
\right.
\label{eq:ansatz}
\end{equation}

Within the interface of width $\xi$, the strain $\epsilon$ linearly
interpolates between $0$ and $\epsilon_0$ (see Fig. 2(a)). The vacancy
field $\phi(x, y, t)$ is obtained from a solution of the diffusion
equation Eq.\ (\ref{eq:phi}), with the initial condition that
$\phi(x,y,0) = \Delta {\bf u} \cdot {\hat {\bf n}}$, where $\Delta {\bf
u}$ can be computed from the variational ansatz Eq.\ (\ref{eq:ansatz})\,;
\begin{eqnarray}
\phi(x,y,t) & = & \Phi(x,y-L/2,t)+\Phi(y,x-L/2,t) + \nonumber \\ 
            &   & \Phi(x,y+L/2,t)+ \Phi(y,x+L/2,t)
\end{eqnarray}
within the interface and $0$ without, where
\begin{eqnarray}
\Phi(a,b,t)  &  =  & \frac{\epsilon_0}{4 \pi \gamma D_{\phi} t}
                     \,\int_{-L/2}^{L/2} 
                dx' \,\int_{-\xi/2}^{\xi/2} dy' \,x'\, \times \nonumber \\
             &     & \exp \left\{-\epsilon_0\,\frac{(a-x')^2+(b-y')^2}{4 
\gamma 
                     D_{\phi} t} \right\} \,.
\label{eq:phisol}
\end{eqnarray}

\myfigure{\epsfysize3.5in\epsfbox{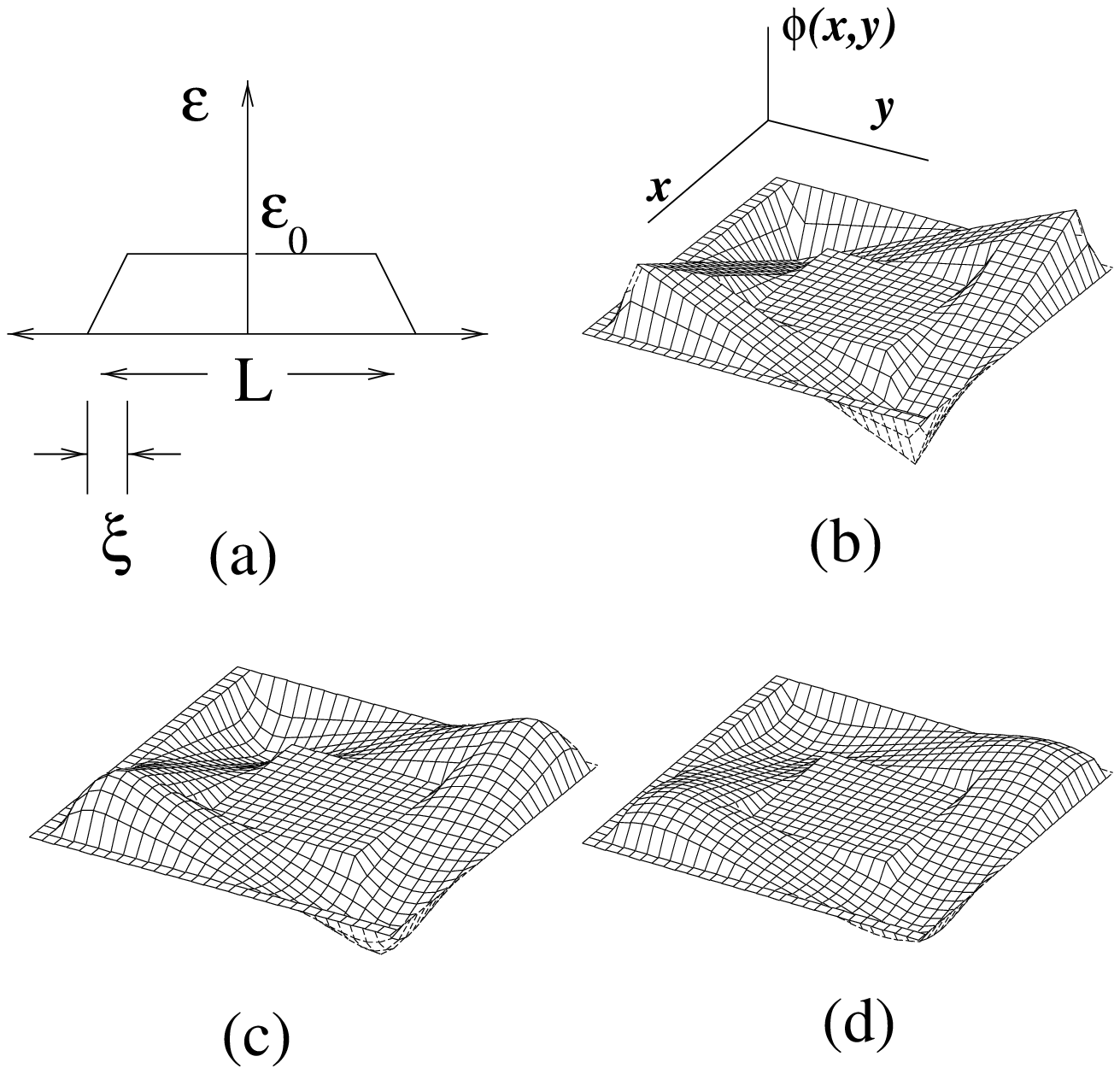}}{\vskip0inFig.\ 2~~(a)
Variational ansatz for $\epsilon$ showing a cross~-section across a
square shaped nucleus of size $L$ and interface width $\xi$. (b)-(d)
3-d plots showing a sequence of $\phi(x,y,t)$ surrounding the nucleus
at different times.}

Figures 2\,(b)-(d) is a sequence of three-dimensional plots of
$\phi(x,y,t)$ at different times. A positive (negative) $\phi$ indicates a
higher interstitial (vacancy) density. The free-energy of a nucleus of
size $L$ at a given time $t$, $E(L,t\,; \xi)$, is obtained by substituting
the variational form for $\epsilon$ and the `field' $\phi(x,y,t)$ in Eq.\
(\ref{eq:FREE2D}). For every $L, t$, we minimize the energy $E(L,t\,;
\xi)$ with respect to the variational parameter $\xi$, to obtain $E(L,t)$.
The energy $\Delta E^*$ and size $L^*$ of the critical nucleus for every
time $t$, is determined by $dE(L,t)/dL\,\vert_{L=L^*} = 0$. The energy of
the critical nucleus decreases with time, and so nucleation of the stable
phase will occur when the time dependent nucleation barrier becomes small
enough. This is obtained by self consistently solving
\begin{equation}
\tau_n = \Gamma^{-1}\, \exp(\Delta E^*(\tau_n))\,.
\label{eq:fpt}
\end{equation}

Once the nucleus exceeds the critical size it grows. Thus the leading
edge of the TTT curves, defined as the time required to form $\delta
\to 0^{+}$ amount of the product at a given temperature of quench,
should be given by Eq.\ (\ref{eq:fpt}). This however does not take
into account the rate at which nuclei are produced. There are
principally two kinds of nucleation modes \cite{RAGH} --- (i) Athermal
and (ii) Isothermal. A number of alloys like the one portrayed in
Fig.\ 1, follow the athermal mode of martensite production.  The
amount of martensite formed is a function only of the temperature of
quench and not the time of holding at that temperature
\cite{ATHERMAL}. A convenient parametrisation is
\begin{equation}
\frac{dN(t)}{dt} = N_0(T) (1-\theta(t-t_0))\,,
\label{eq:athermal}
\end{equation}
where $t_0 \ll \tau_n$ and $N_0(T)$ increases with decreasing $T$ below
$M_s$. It is clear that as long as $t_0 \ll \tau_n$, all grains are 
essentially nucleated at once, and the leading edge
of the TTT curves is given by $\tau_n$ (solid line in Figure 3).  The
curves in Figure 3, have been derived using the quoted values for the
parameters $D_{\infty}$ and $A$ for Fe alloys \cite{LONG}. The modulus
$\gamma$ is fixed by the condition that the intercept $a_{M_s}(\gamma)
= M_s/4\,T_s$. For Fe alloys, $M_s = 100 - 200^{\circ}$C and $T_s =
900^{\circ}$C, leading to a $\gamma = 0.1$. The time axis is in units
of $\Gamma^{-1}$ taken to be $10^{-6}$s. All these parameters can be varied
over a wide range without changing the qualitative nature of the results.

For low undercooling (small $1/4 - a$), the critical barrier height is large.
The $\phi$ field relaxes quickly, and so $\tau_n$ asymptotes the 
$D_{\phi}=\infty$ curve (dotted line in Fig.\ 3). Once the critical nucleus 
is formed, it grows as a ferrite \cite{PRL}.
For larger undercooling, $D_{\phi}$ decreases reaching a vanishingly small
value for  $a \le a_{M_s}$. This part of $\tau_n(a)$ 
asymptotes the $D_{\phi}=0$ curve,
when the $\phi$ field remains frozen at the parent/product interface. As
shown in Ref.\ \cite{PRL}, the nucleus twins in the direction of motion 
resulting in a martensite. 

\vspace{0cm}
\myfigure{\epsfysize2in\epsfbox{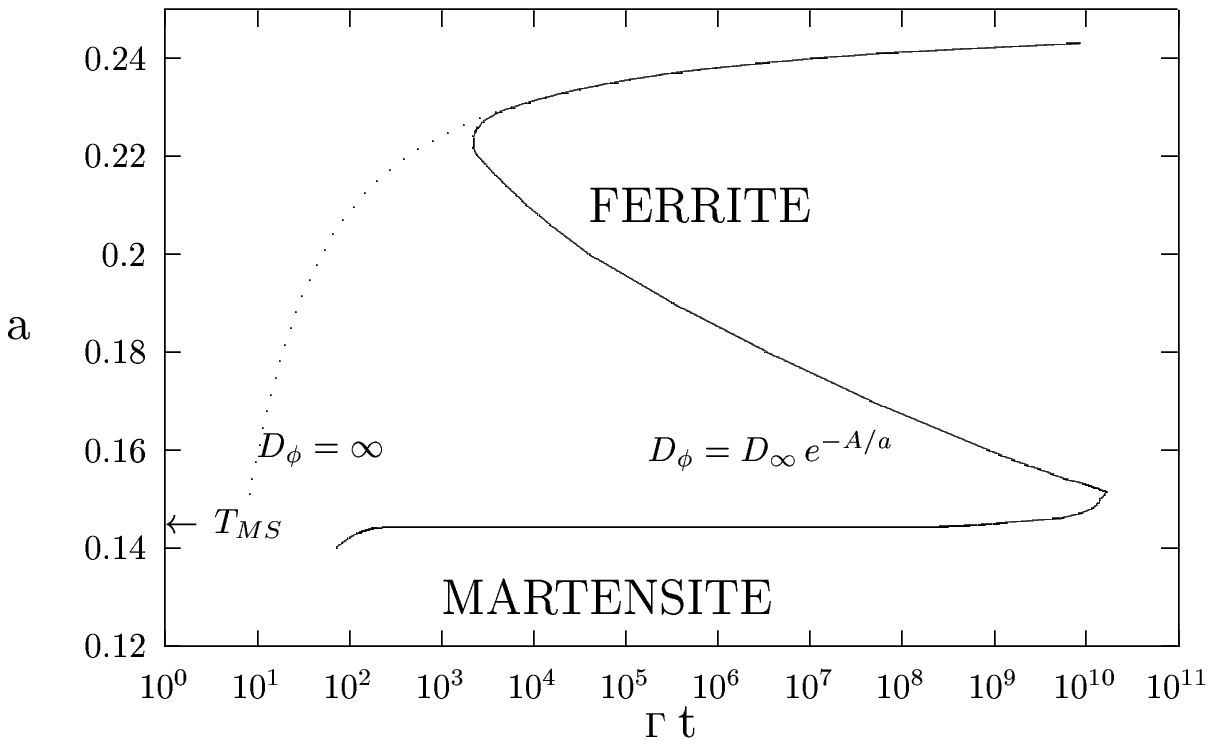}}{\vskip0inFig.\ 3~~Calculated
leading edge of the TTT curves showing ferrite and athermal martensite
regions (solid line). The values for $D_{\infty}$ and $A$ in dimensionless
units have been translated from the quoted values for Fe (see text) ---
$D_{\infty} = 4 \times 10^{14}$ and $A = 8$. The arrow indicates
the martensite-start temperature $a_{M_s}$ in our units. The dashed curve
shows the first passage time for $D_{\phi}=\infty$}

Our calculation thus successfully reproduces the two distinctive features
of athermal martensites --- the horizontal martensite transformation
curves, and a well defined martensite-start temperature $M_s$ which is
{\it independent} of $D_{\infty}$ and decreases with increasing $\gamma$. 

On the other hand in isothermal martensites, the nuclei are thermally
generated, and so $t_0$ is large compared to $\tau_n$. In this situation,
the leading edge of the TTT curve does not have a precise meaning. We can
nevertheless define the leading edge from the product $(dN/dt)
\,\tau_n^{-1}$. Being produced by a thermal activation process, $dN/dt$
decreases with increased undercooling. However as stated above, the
critical barrier height and hence $\tau_n$ decreases with increased
undercooling. This produces the charateristic $C$-shaped TTT curves of the
isothermal martensite \cite{RAGH}. The qualitative features of the leading
edge do not change if the isothermal nucleation is autocatalytic. 

The other curves in the TTT family (Fig.\ 1) correspond to the
transformation of a fixed amount of the product. This requires a detailed
knowledge of the dynamics of patterning generated by the nucleation and
growth of grains \cite{KS}, together with the shape and size of the
individual grains. 

In summary, we have derived the essential features of the leading edge of
the time-transformation-temperature curves for both athermal and
isothermal martensites, within the first-passage time formalism. This
makes use of a recently developed \cite{PRL} mode-coupling theory for the
morphology and nucleation kinetics of the equilibrium ferrite and twinned
martensites. Such transformation curves have thus far been obtained
empirically, and to the best of our knowledge have never been derived. 

Most martensites contain alloying elements which occur as interstitial or
substitutional impurities. The solubility of these impurities may be
different in the parent and product phases (eg., interstitial carbon in
steel). This would induce large scale diffusion of impurities as the
transformation proceeds (eg., carbon diffuses away from the ferritic
product). TTT diagrams of such alloy steels often show a secondary bulge
together with the formation of an intermediate structure called {\it
bainite}. These issues may be understood within our formalism by coupling
the density of impurities to the strain $\epsilon$.  We investigate the
role of alloying in a forthcoming publication. 

We thank K.\ P.\ N.\ Murthy for a discussion on time-dependent nucleation 
barriers.


\begin{references}

\vspace{-1cm}

\bibitem{NISHROIT}
A.\ Roitburd, in {\it Solid State Physics}, ed.\ Seitz and Turnbull
(Academic Press, NY, 1958)\,; Z.\ Nishiyama, {\it Martensitic
Transformation} (Academic Press, NY, 1978)\,; A.\ G.\ Kachaturyan, {\it
Theory of Structural Transformations in Solids} (Wiley, NY, 1983). 

\bibitem{KRUM}
G.\ R.\ Barsch and J.\ A.\ Krumhansl, Phys.\ Rev.\ Lett.\ {\bf 37}, 9328
(1974); G.\ R.\ Barsch, B.\ Horovitz and J.\ A.\ Krumhansl, Phys.\ Rev.\
Lett.\ {\bf 59}, 1251 (1987). 

\bibitem{GOOD}
G.\ S.\ Bales and R.\ J.\ Gooding, Phys.\ Rev.\ Lett.\ {\bf 67}, 3412
(1991)\,; A.\ C.\ E.\ Reed and R.\ J.\ Gooding, Phys.\ Rev.\ {\bf B50},
3588 (1994)\,; B.\ P.\ van Zyl and R.\ J.\ Gooding,
http://xxx.lanl.gov/archive/cond-mat/9602109. 

\bibitem{MH}
{\it Metals Handbook}, 9$^{th}$ Edition, Vol. 4 (ASM, Ohio, 1981).

\bibitem{PRL}
M.\ Rao and S.\ Sengupta, Phys.\ Rev.\ Lett.\ {\bf 78}, 2168 (1997).

\bibitem{LL}
L.\ D.\ Landau and E.\ M.\ Lifshitz, in {\it Theory of Elasticity} (Pergamon
Press, 1989).

\bibitem{NOTE1}
The product might prefer ro generate dislocations at the parent-product
interface (producing internal slip bands).

\bibitem{SMITH}
{\it Smithells Metals Reference Book}, 7$^{th}$ edition, eds.\ E.\ A.\
Brandes and G.\ B.\ Brook (Butterworth~-Heinemann, Oxford, 1992).

\bibitem{AS}
G.\ S.\ Agarwal and S.\ R.\ Shenoy, Phys.\ Rev.\ A {\bf 23}, 2719
(1981)\,; S.\ Chandrasekhar in {\it Selected Papers on Noise and
Stochastic Processes}, ed.\ N.\ Wax (Dover, NY, 1957). 

\bibitem{RAGH}
V.\ Raghavan, in {\it Martensite}, eds.\ G.\ B.\ Olson and W.\ S.\ 
Owen (ASM International, The Materials Information Society, 1992).

\bibitem{ATHERMAL}
The microscopic explanation of such a nucleation process is unclear,
though some experiments (V.\ Raghavan and M.\ Cohen, Acta Metall., {\bf
20}, 1251 (1972)) suggest a thermal origin. 

\bibitem{LONG}
The qualitative features of the leading edge of the TTT curves, depend only
on gross features of the time dependence of the barrier height. In 
particular they are insensitive to the nature of the order parameter, 
dimensionality, shape of the nucleus etc. This aspect allows us to use the
measured parameters for Fe alloys (a 3-dim FCC $\to$ BCC transition) in our 
calculation.     

\bibitem{KS}
M.\ Rao, S.\ Sengupta and H.\ K.\ Sahu, Phys.\ Rev.\ Lett.\ {\bf 75}, 2164
(1995)\,; E.\ Ben-Naim and P.\ Krapivsky, Phys.\ Rev.\ Lett.\ {\bf 76},
3234 (1996)\,; M.\ Rao and S.\ Sengupta, Phys.\ Rev.\ Lett.\ {\bf 76},
3235 (1996). 

\end{references}
\end{document}